\documentclass[12pt]{iopart}

\usepackage{graphicx}% Include figure files
\usepackage{braket}
\usepackage{array}
\usepackage{color} 
\usepackage{bm}
\usepackage{braket}
\begin{document}

\title{The Berry curvature of the Bogoliubov quasiparticle Bloch states in the unconventional superconductor Sr$_2$RuO$_4$}

\author{Martin Gradhand and James F. Annett}

\address{H . H. Wills Physics Laboratory, University of Bristol, Tyndall Ave, BS8-1TL, UK}
\ead{m.gradhand@bristol.ac.uk}

\begin{abstract}
We will extend the concept of electron band Berry curvatures to superconducting materials. We show that this can be defined for the Bogoliubov-de Gennes equation describing the superconducting state in a periodic crystal. In addition, the concept is exploited to understand the driving mechanism for the optical Kerr effect in time reversal symmetry breaking superconductors. Finally, we establish a sum rule analogue to the normal state Hall sum rule making quantitative contact between the imaginary part of the optical conductivity and the Berry curvature. The general theory will be applied and tested against the drosophila of the p-wave paired materials Sr$_2$RuO$_4$.
\end{abstract}

\maketitle

\section{Introduction}
The Berry curvature of Bloch electrons has recently turned out to be a powerful and unifying concept in the linear response theory of electrons in periodic crystals under applied electromagnetic fields.\cite{Berry_1984, Xiao2010, Gradhand2012b} The methodology was not only successful in providing a deep insight into the physical mechanisms leading to the anomalous Hall effect, spin Hall effect, electric polarization, orbital magnetization, and circular dichroism in metals and insulators but was also useful to provide a numerical tool to address problems which would be extremely challenging otherwise.\cite{Berry_1984}
We will extend this concept to the Bogoliubov-de Gennes equation in $\bf{k}$ space, describing the superconducting state and its quasiparticle excitations. The basic definitions will be outlined in the following section. Using the Hall sum rule or equivalently the Kramers-Kronig transformation we make contact between the imaginary part of the frequency dependent optical conductivity and the Berry curvature of the Bogoliubov-de Gennes equation. In section 4 we will apply these definitions and derivations to the superconducting state of Sr$_2$RuO$_4$. Here we discuss typical features of the Berry curvature in the superconducting state and verify the established Hall sum rule numerically. This leads to a further microscopic understanding of the intrinsic mechanism leading to the optical Kerr effect in p-wave superconductors. \cite{Yip1992,Read2000,Yakovenko2007,Mineev2007,Lutchyn2008,Roy2008,Taylor2012,Wysokinski2012} Furthermore, we will include spin-orbit coupling and analyse its influence on the Berry curvature and optical properties.

\section{The Berry curvatures of the Bogoliubov-de Gennes equation}

According to M. Berry \cite{Berry_1984} the curvature arises from the adiabatic approximation to the time dependent solution of a Hamiltonian including a parameter. Most importantly, it was shown that the Berry curvature is a gauge invariant quantity which, as soon as it can be defined, will have implications for physical observables. \cite{Berry_1984,Xiao2010,Niu99} In the case of periodic crystals and the related Bloch states $\Psi_{n{\bf k}}({\bf r})=e^{i{\bf k r}}u_{n{\bf k}}({\bf r})$, the operator $\hat{H}_{\bf k}({\bf r})=e^{-i{\bf k r}}\hat{H}({\bf r})e^{i{\bf k r}}$ generating the periodic part of the Bloch function $u_{n{\bf k}}({\bf r})$ is parameter dependent. \cite{Xiao2010,Niu99,Gradhand2012b} The relevant parameter is the Bloch wave vector $\bf{k}$ and the solutions of the eigenvalue equations, related to different wave vectors, are independent from  another. Here $n$ is the band index. The success of the concept of geometrical phases and Berry curvatures in the context of condensed matter theory relies on that connection. Also derived from the adiabatic approximations it resembles other approaches and is often elucidating as a unifying concept.~\cite{Xiao2010,Gradhand2012b}

Turning our attention to the superconducting case our starting point is the Bogoliubov-de Gennes (BdG) equation for the wave function $\braket{{\bf r}|\Psi_{n{\bf k}}}=\left(
u_{n}({\bf r}),v_{n}({\bf r})
\right)^T $ considering a local gap function \cite{Ketterson1999}
\begin{equation}\label{Eq:BdG_Bloch}
\left(
\begin{array}{cc}
\hat{H}({\bf r})& \hat{\Delta}({\bf r})\\
\hat{\Delta}^{\dagger}({\bf r})&-\hat{H}^{\ast}({\bf r})
\end{array}
\right)\left(
\begin{array}{c}
u_{n}({\bf r})\\
v_{n}({\bf r})
\end{array}
\right)
=E_{n}
\left(
\begin{array}{c}
u_{n}({\bf r})\\
v_{n}({\bf r})
\end{array}
\right)\ ,
\end{equation}
where $n$ labels all eigenstates. For convenience we suppressed all spin indices in the above equation as we will do in the following to focus on the main new feature namely coupling electrons and holes and its impact on the Berry curvature. Nevertheless, it should be stated that for the consideration of the unconventional superconducting state of Sr$_2$RuO$_4$ the spin degree of freedom has to be included. Following from this all entries in matrices have to be considered as two by two matrices in spin space and $u_{n}({\bf r})$ and $v_{n}({\bf r})$ are spinors each. All derivations will hold but the degeneracy in spin space leads to implications further discussed in section 4. For a periodic crystal we can separate the Bloch phase factor from the lattice periodic part of the wave function as in the normal state \cite{Ketterson1999}
$\braket{{\bf r}|\Psi_{n{\bf k}}}=e^{i{\bf kr}}\left(u_{n{\bf k}}({\bf r}),v_{n{\bf k}}({\bf r})
\right)^T $ 
and retrieve an equation for the periodic wave function within one unit cell (UC)\cite{Ketterson1999}
\begin{equation}\label{Eq:BdG}
\left(
\begin{array}{cc}
\hat{H}_{\bf k}({\bf r})& \hat{\Delta}({\bf r})\\
\hat{\Delta}^{\dagger}({\bf r})&-\hat{H}_{-{\bf k}}^{\ast}({\bf r})
\end{array}
\right)\left(
\begin{array}{c}
u_{n{\bf k}}({\bf r})\\
v_{n{\bf k}}({\bf r})
\end{array}
\right)
=E_{n{\bf k}}
\left(
\begin{array}{c}
u_{n{\bf k}}({\bf r})\\
v_{n{\bf k}}({\bf r})
\end{array}
\right)\ .
\end{equation}
Here, the ${\bf k}$-dependent lattice periodic normal state Hamiltonian $\hat{H}_{\bf k}=e^{-i{\bf kr}}\hat{H}({\bf r)} e^{i{\bf kr}}$ appears on the diagonals and the ${\bf k}$ independent local gap function $\hat{\Delta}({\bf r})$ is connecting electron and hole like states on the off-diagonal part of the operator. The local gap functions are ${\bf k}$ independent on that level since they commute with the Bloch phase factor which cancels from the left and right side of the equation. It should be noted that the gap function for a particular tight-binding model might be ${\bf k}$ dependent but this needs to be separated from the ${\bf k}$ dependence induced moving from the equation for the Bloch function to the expression for the periodic part only. Equation (\ref{Eq:BdG}) is clearly an eigenvalue problem including the parameter ${\bf k}$ with a similar structure as the eigenvalue equation of the normal state. 

Following this, we can define the Berry curvature for the superconducting state keeping in mind that this gauge invariant quantity should have physical implications we will derive in the following. 
The formal definition of the Berry curvature of Bloch states is~\cite{Berry_1984,Xiao2010,Gradhand2012b,Ghosh2010} 
\begin{equation}
\boldsymbol{\Omega}_n({\bf k})=i\boldsymbol{\nabla}_{\bf k} \times \int \limits_{\rm UC} \rm{d}^3\rm{r}\  \Phi^{\ast}_{n{\bf k}}({\bf r})\boldsymbol{\nabla}_{\bf k}\Phi_{n{\bf k}}({\bf r})=i\boldsymbol{\nabla}_{\bf k} \times \left(\Phi_{n{\bf k}}({\bf r}),\boldsymbol{\nabla}_{\bf k}\Phi_{n{\bf k}}({\bf r})\right)\ \rm{,}
\end{equation}
where $(\cdot,\cdot)$ is a shorthand notation for the inner product of the periodic part of the Bloch function defined as the real space integral over the unit cell (UC) only. Using the standard procedure \cite{Berry_1984,Gradhand2012b} to express the ${\bf k}$ derivative of the wave function in terms of the ${\bf k}$ derivative of the operator we rewrite the expression as
\begin{equation}\label{Eq.:Berry}
\boldsymbol{\Omega}_n({\bf k})=\sum\limits_{m\neq n} \frac{\left(\Phi_{n{\bf k}}({\bf r}),\boldsymbol{\nabla}_{\bf k} {\hat M}_{\bf k}({\bf r})\Phi^{\ast}_{m{\bf k}}({\bf r})\right)\times  \left(\Phi_{m{\bf k}}({\bf r}), \boldsymbol{\nabla}_{\bf k} {\hat M}_{\bf k}({\bf r})\ \Phi_{n{\bf k}}({\bf r})\right)} {\left(E_m({\bf k})-E_n({\bf k})\right)^2}\ \rm{.}
\end{equation}
Where ${\hat M}_{\bf k}({\bf r})$ is the ${\bf k}$ dependent operator of the Bogoliubov-de Gennes equation,
\begin{equation}
{\hat M}_{\bf k}({\bf r})=\left(
\begin{array}{cc}
\hat{H}_{\bf k}({\bf r})& \hat{\Delta}({\bf r})\\
\hat{\Delta}^{\dagger}({\bf r})&-\hat{H}_{-{\bf k}}^{\ast}({\bf r})
\end{array}
\right)\ {\rm .}
\end{equation}
Its eigenvectors are the periodic part of the Bogoliubov Bloch functions 
\begin{equation}
\Phi_{n{\bf k}}({\bf r})=\left(
\begin{array}{c}
u_{n{\bf k}}({\bf r})\\
v_{n{\bf k}}({\bf r})
\end{array}
\right)\ {\rm ,}
\end{equation}
and the energies $E_n({\bf k})$ are the corresponding eigenvalues. Equation~(\ref{Eq.:Berry}) gives an expression for the Berry curvature of the Bogoliubov-de Gennes equation. This result is very similar to the discussion of Ghosh et al. \cite{Ghosh2010} who introduced a Berry curvature for the BdG-equation as well. The minor differences relate to the fact that we introduce a band resolved Berry curvature, $\Omega_n({\bf k})$ , in contrast to the sum over all occupied states as performed by them. \cite{Ghosh2010} In the following we will simplify the ${\bf k}$ derivative of the BdG operator and make contact between the defined curvature and physical observables such as the optical conductivity. 
Due to the ${\bf k}$ independence of the gap function and assuming a real normal state Hamiltonian the derivative of the ${\bf k}$-dependent operator can be simplified to
\begin{equation}
\boldsymbol{\nabla}_{\bf k} {\hat M}_{\bf k}({\bf r})=\left(
\begin{array}{cc}
\boldsymbol{\nabla}_{\bf k} \hat{H}_{\bf k}({\bf r})& 0\\
0 &\boldsymbol{\nabla}_{\bf k} \hat{H}_{{\bf k}}({\bf r})
\end{array}
\right)\ {\rm .}
\end{equation}
Furthermore, we used the identity $\boldsymbol{\nabla}_{\bf k} \hat{H}_{{\bf k}}({\bf r})=-\boldsymbol{\nabla}_{\bf k} \hat{H}_{-{\bf k}}({\bf r})$.
Including on-site spin-orbit coupling would not affect the above result since it does not introduce any complex ${\bf k}$ dependence to the Hamiltonian. To that end we have defined the Berry curvature. However, its impact on physical properties of the superconducting state has to be shown separately. Nevertheless, we can expect its influence due to its gauge invariant structure.

\section{The Hall sum rule}

From the normal state it is well known that the Berry curvature of the ground state can be related to the frequency dependent optical conductivity via the Hall sum rule, or equivalently the Kramers-Kronig transformation. It was shown by I. Souza and D. Vanderbilt that in this case the relation \cite{Souza2008}
\begin{equation}\label{Eq.:Sum_norm}
\int\limits_0^\infty{\rm d} \omega\ \frac{{\rm Im}\left(\sigma_{xy}(\omega)\right)}{\omega}=-\frac{e^2\pi}{2\hbar}\sum_n\int\frac{{\rm d}^3 {\rm k}}{(2\pi)^3}\Omega^z_n({\bf k})\ f(E_n({\bf k}))
\end{equation}
holds, where the sum on the right hand side runs over occupied states only given by the Fermi-Dirac distribution $f(E_n({\bf k}))$. The superscript $z$ indicates the $z$ component of the Berry curvature vector.

According to linear response theory the imaginary part of the optical conductivity in the superconducting state can be expressed as \cite{Capelle1997,Capelle1998,Gradhand2013} 
\begin{eqnarray}\label{Eq.:Imsxy}
{\rm Im}\left[\sigma_{xy}(\omega)\right]=\frac{\pi e^2}{2\omega V\hbar^2}\sum\limits_{n,m, {\bf k}}f(E_{n{\bf k}})\left[1-f(E_{m{\bf k}})\right]\delta(E_{m{\bf k}}-E_{n{\bf k}}-\hbar\omega)\times\nonumber\\{\rm Im}\left[\braket{ {m {\bf k}} |\hat{H}^x_{\bf k}|{n {\bf k}}}\braket{ {n {\bf k}}|\hat{H}^y_{\bf k}|{m {\bf k}}}-\braket{ {m {\bf k}} |\hat{H}^y_{\bf k}|{n {\bf k}}}\braket{ {n {\bf k}} |\hat{H}^x_{\bf k}|{m {\bf k}}}\right]
\end{eqnarray} 
where we used 
$\braket{{\bf r}|{n{\bf k}}}=\Phi_{n{\bf k}}({\bf r})=\left(
u_{n{\bf k}}({\bf r}),v_{n{\bf k}}({\bf r})
\right)^T $
for the periodic part of the Bloch function within the BdG equation and the shorthand notation 
\begin{equation}\label{Eq:Hint3}
\braket{{{\bf r}}|\hat{H}^{x}_{\bf k}|{{\bf r}^\prime}}=\delta({\bf r}-{\bf r}^\prime)
\left(
\begin{array}{cc}
\partial_{k_{x}}\hat{H}_{\bf k}({\bf r})&0\\
0&\partial_{k_{x}}\hat{H}_{\bf k}({\bf r})
\end{array}\right)\ {\rm .}\\
\end{equation}
for the interaction operator and similarly for $\hat{H}^{y}_{\bf k}$ .
Performing the integral over all frequencies yields
\begin{eqnarray}
\int\limits_0^\infty{\rm d} \omega\ \frac{{\rm Im}\left(\sigma_{xy}(\omega)\right)}{\omega}=
\frac{\pi e^2}{2V\hbar}\sum\limits_{n,m, {\bf k}}f(E_{n{\bf k}})\left[1-f(E_{m{\bf k}})\right]\times\nonumber\\
\frac{{\rm Im}\left[\braket{ {m {\bf k}} |H^x_{\bf k}|{n {\bf k}}}\braket{ {n {\bf k}}|H^y_{\bf k}|{m {\bf k}}}-\braket{ {m {\bf k}} |H^y_{\bf k}|{n {\bf k}}}\braket{ {n {\bf k}} |H^x_{\bf k}|{m {\bf k}}}\right]}{\left(E_m({\bf k})-E_n({\bf k})\right)^2}\ {\rm ,}
\end{eqnarray}
which has almost the desired form of Eq.~(\ref{Eq.:Berry}) to be rewritten in terms of the Berry curvature defined above. The main difference comes from the presence of $1-f(E_{m{\bf k}})$ restricting the sum over $m$ to unoccupied states only. However, it was shown previously that all contributions from occupied states within the sum over m and n vanishes. \cite{Souza2008} To be more explicit we have the condition
\begin{equation}
\sum\limits_{n,m, {\bf k}}f(E_{n{\bf k}})f(E_{m{\bf k}}) 
\frac{{\rm Im}\left[\braket{ {n {\bf k}} |\boldsymbol{\nabla}_{\bf k} {\hat M}_{\bf k}|{m {\bf k}}}\times\braket{ {m {\bf k}}|\boldsymbol{\nabla}_{\bf k} {\hat M}_{\bf k}|{n {\bf k}}}\right]}{\left(E_m({\bf k})-E_n({\bf k})\right)^2} = 0 \ {\rm ,}
\end{equation}
and the sum over the Berry curvature over all occupied bands can be rewritten as
\begin{eqnarray}
\sum\limits_{n {\bf k}}\boldsymbol{\Omega}_n({\bf k})f(E_{n{\bf k}})=
\sum\limits_{n,m, {\bf k}}f(E_{n{\bf k}})\left[1-f(E_{m{\bf k}})\right]\times \nonumber\\
\frac{{\rm Im} \left[\braket{ {n {\bf k}} |\boldsymbol{\nabla}_{\bf k} {\hat M}_{\bf k}|{m {\bf k}}}\times\braket{ {m {\bf k}}|\boldsymbol{\nabla}_{\bf k} {\hat M}_{\bf k}|{n {\bf k}}}\right]}{\left(E_m({\bf k})-E_n({\bf k})\right)^2}\ {\rm .} 
\end{eqnarray}
This yields immediately the desired Hall sum rule relating the integral over the optical conductivity to the sum over all occupied bands of the Berry curvature 
\begin{equation}\label{Eq.:sum_rule_super}
\int\limits_0^\infty{\rm d} \omega\ \frac{{\rm Im}\left(\sigma_{xy}(\omega)\right)}{\omega}=-
\frac{\pi e^2}{2\hbar}\sum\limits_{n}\int\frac{{\rm d}^3{\rm k}}{(2\pi)^3}\Omega_n^z({\bf k})f(E_{n{\bf k}})
\ {\rm ,}
\end{equation}
where we used the introduced definitions and the replacement of the sum over ${\bf k}$ states by the integral $\sum\limits_{\bf k}\rightarrow \frac{V}{(2\pi)^3}\int {\rm d}^3{\rm k}.$ This is the equivalent to the normal state expression of Eq.~(\ref{Eq.:Sum_norm}) for the superconducting state exploiting the Berry curvature expression as defined in Eq.~(\ref{Eq.:Berry}). It has to be pointed out that also the expression looks exactly the same as in the normal state the physics is quite different. All quantities in Eq.~(\ref{Eq.:sum_rule_super}) refer to their expressions in the superconducting state, ie. Eqs.~(\ref{Eq.:Berry}) and (\ref{Eq.:Imsxy}). These are defined via the Bogoliubov quasiparticles and are distinct from the normal state analogues. Nevertheless, the power of this generalization of the Berry curvature to the superconducting state is that the tools and the understanding from the normal state can be transferred to superconductors. The Berry curvatures as unifying concept can be exploited in yet another context.

Finally, we would like to close this section by pointing out that due to Kramers-Kronig transformation the integral of Eq.~(\ref{Eq.:sum_rule_super}) is related to the zero frequency real part of the optical conductivity and we can connect this quantity via
\begin{equation}\label{Eq.:KKT}
{\rm Re}\left(\sigma_{xy}(\omega=0)\right)=-
\frac{ e^2}{\hbar}\sum\limits_{n}\int\frac{{\rm d}^3{\rm k}}{(2\pi)^3}\Omega_n^z({\bf k})f(E_{n{\bf k}})
\ {\rm ,}
\end{equation}
to the Berry curvature of the BdG quasiparticles.
 
\section{Application to the unconventional superconductor Sr$_2$RuO$_4$}

In the following we will analyse the Berry curvature (Eq.~(4)) and the related Hall sum rule (Eq.~(14)) in the superonducting state of Sr$_2$RuO$_4$ in detail. The calculations will rely on an empirical tight-binding model describing the p-wave paired state of Sr$_2$RuO$_4$ in quantitative agreement to several experimental observations.\cite{Mackenzie1996,Mackenzie1998,Nishizaki2000,Bergemann2000} The underlying model was explained in previous publications \cite{Gradhand2013,Annett2002} and is not the focus of the current work. Here, we only state the most important feature of this model relevant for the discussion of the Berry curvature. The model is restricted to the 3 Ru-d (d$_{xy}$, d$_{xz}$, d$_{yz}$) orbitals present a the Fermi level. The hopping parameters are chosen to reproduce the experimentally found Fermi surface and bandwidths. On top of that the BdG  is solved self-consistently for unconventional p-wave pairing between opposite spins relying on a minimal set of two interaction parameters to reproduce the experimentally found critical temperature, the specific heat, and the superfluid density. \cite{Annett2002,Annett2003} The most important feature of the model is the time reversal symmetry breaking induced by the superconducting state.\cite{Gradhand2013,Annett2002} The special symmetry of the considered gap function and its time reversal symmetry breaking is a possible mechanism to explain various experimental observations \cite{Nishizaki2000,Izawa2001,Annett2002} most importantly the finite Kerr signal at optical frequencies \cite{Gradhand2013,Xia2006,Kapitulnik2009,Wysokinski2012}. 

All the results shown in the following are based on the self-consistent solution of the BdG equation for a particular set of parameters details of which are presented in Ref.~\cite{Gradhand2013}. Let us start with the quasiparticle bandstructure as shown in Fig.~\ref{fig:band}. All together we have 6 two-fold degenerate bands, 3 of which are positive and three with negative energy eigenvalues. Positive and negative solutions are related by symmetry and the system is over determined. For that reason at $T=0$ it is sufficient to consider the negative solutions only, which we will do in the following.
\newlength{\LL} \LL 0.8\linewidth
\begin{figure}[ht]
\includegraphics[width=\LL]{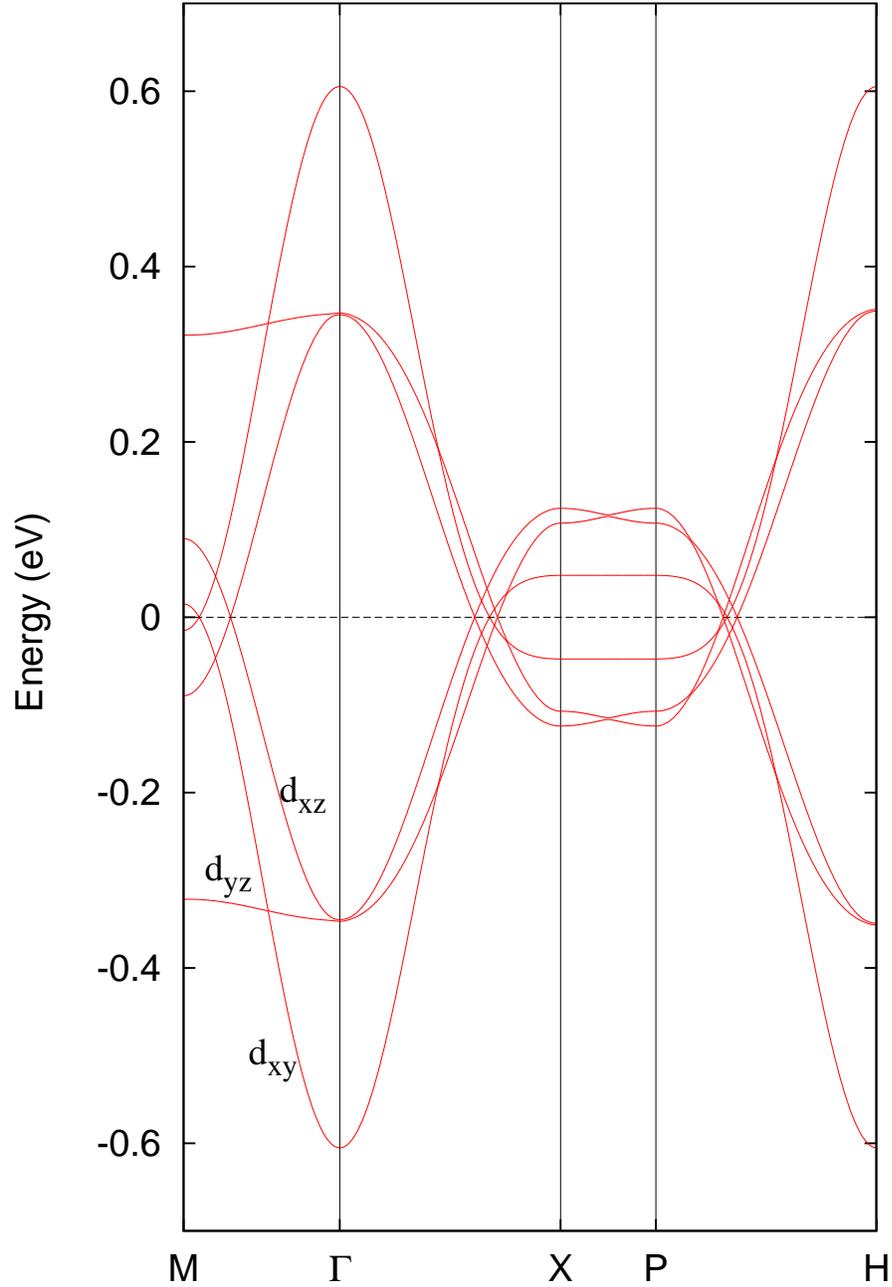}
\caption{Quasi particle bandstructure of the considered three band model for the superconducting phase of Sr$_2$RuO$_4$ along high symmetry lines. \label{fig:band}}
\end{figure}
Along the symmetry lines we find several symmetry induced and accidental crossings as well as avoided crossings. It is well known that the Berry curvature is induced by the close vicinity of neighbouring bands and accidental degeneracies or avoided crossings are a particular source of Berry curvatures \cite{Berry_1984,Xiao2010,Gradhand2012b,Wang2006}. At $E=0$ the superconducting state induces a gap which is not visible on the scale shown in Fig.~\ref{fig:band} but its structure was discussed in detail in previous publications \cite{Gradhand2013,Annett2002, Annett2003,Wysokinski2003}. 

Without spin-orbit coupling and neglecting the gap function the Berry curvature of the system would be zero at all ${\bf k}$ points and we will investigate in the following how both will create distinct features for the Berry curvature along the symmetry lines shown in Fig.~\ref{fig:band}. Here, we would like to highlight one further point. As outlined following Eq.~(1) we implicitly considered all entries in matrices being two by two matrices or vector components being spinors in spin space, respectively. This is essential to describe the unconventional superconducting state of Sr$_2$RuO$_4$ and leads to a twofold degeneracy at all ${\bf k}$ points. Due to this degeneracy of the bands we encounter a slight complication of the matter since instead of an Abelian Berry curvature, where each component of the vector is a scalar, we have to consider the non-Abelian  Berry curvature, where each component is a matrix,. \cite{Gradhand2012b,Culcer2005,Shindou2005,Wilczek1984} Each entry within the matrix is gauge dependent but quantities like the trace $Tr[\Omega^z_n({\bf k})]=\left(\Omega^z_n({\bf k}\right)_{11}+\left(\Omega^z_n({\bf k}\right)_{22})$ are gauge independent.\cite{Gradhand2011} For the normal state it is well known that the trace of the Berry curvature of time and space inversion symmetrical systems is zero due to the fact of a zero anomalous (charge) Hall effect. \cite{Shindou2005} Nevertheless, including spin-orbit coupling the individual entries of the matrix become finite and $Tr[\sigma_z\Omega^z_n({\bf k})]=\left(\Omega^z_n({\bf k}\right)_{11}-\left(\Omega^z_n({\bf k}\right)_{22})$ is non zero. Here $\sigma_z$ is a Pauli matrix and the calculated quantity is responsible for the intrinsic spin Hall effect. 

The same symmetry arguments hold for the superconducting state, where however the gap function breaks time reversal symmetry in a special way preserving the two fold degeneracy of the bands. This leads to the result that neglecting spin-orbit coupling leads to $Tr[\sigma_z\Omega^z_n({\bf k})]=0$ for all k-points. Whereas neglecting the influence of the gap but considering spin-orbit coupling would result into $Tr[\Omega^z_n({\bf k})]=0$. Considering both simultaneously leads to the interesting situation of finite contributions to both traces similar to the situation of a normal ferromagnet where we find a finite anomalous (charge) Hall effect but on top of it a spin Hall effect as well. Although in the case of a collinear magnet such as Fe we lift the degeneracy of the bands and the Berry curvature becomes Abelian again.

Let us consider first the case of $Sr_2RuO_4$ without spin-orbit coupling to focus on the effect of the gap function only. 
 \begin{figure}[h]
\includegraphics[width=\LL]{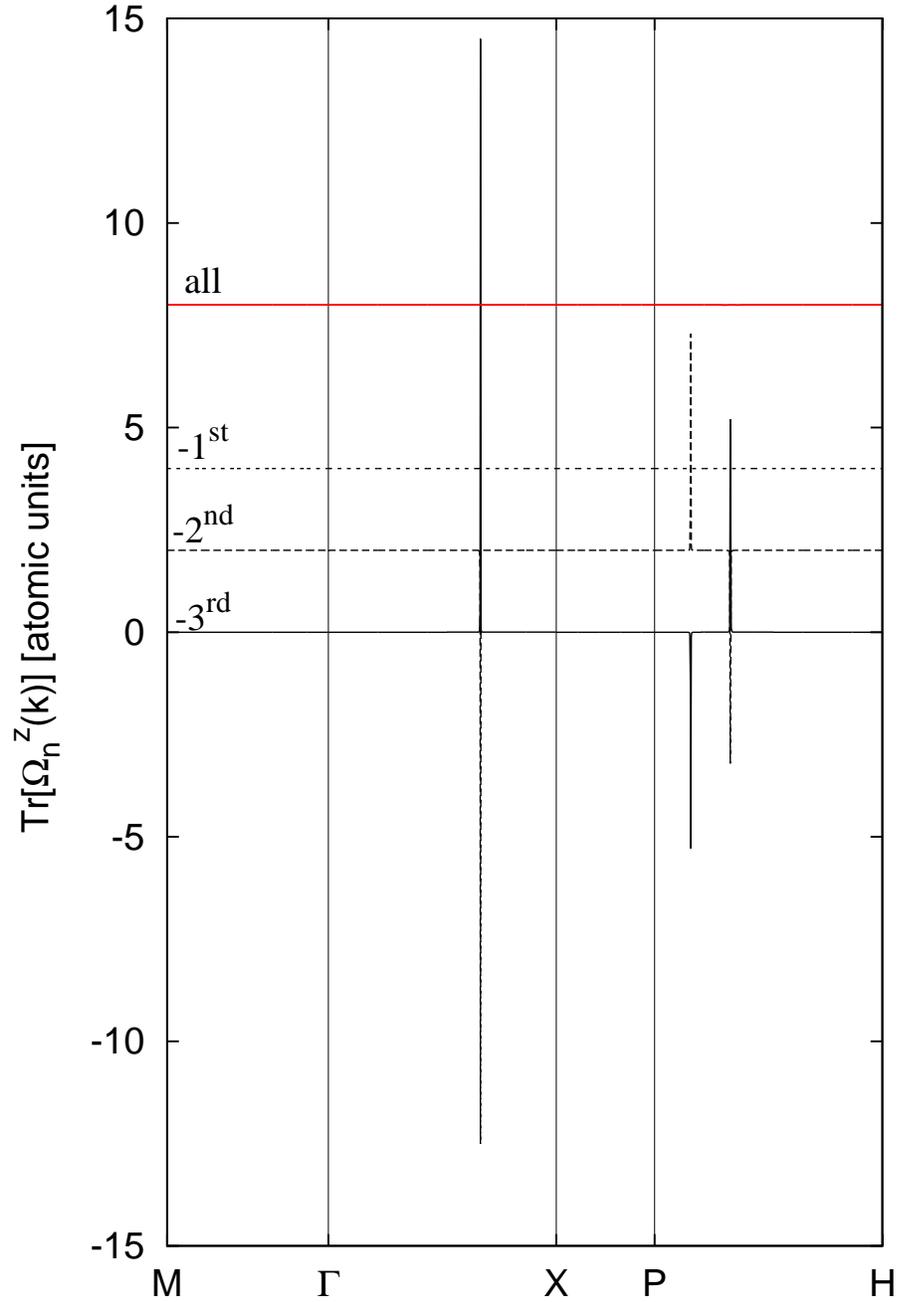}
\caption{The trace of the 2-dim. non-Abelian Berry curvature for each degenerate band with negative eigenvalue in Sr$_2$RuO$_4$ along high symmetry lines neglecting spin-orbit coupling. For visibility the curves are shifted by a constant offset and the upper curve is the sum over all bands.\label{fig:Berry_noSOC_full}}
\end{figure}
In Fig.~\ref{fig:Berry_noSOC_full} we show the band resolved Berry curvature of the three lowest energy bands along high symmetry lines. The curves for each band are shifted by a constant amount for a better visibility and the upper (red) curve is the sum over all bands. On that scale only a very few features are visible which essentially stem from near degeneracies between the lowest energy bands along the chosen path in ${\bf k}$ space. Especially, we point out that these features are relatively far away from the actual superconducting gap, have opposite sign within the two bands and enters the sum in such a way that their contribution vanishes for any physical observable. Furthermore, the size of these features is probably strongly enhanced by the simplicity of the underlying tight-binding model with a small number of orbital overlaps leading to very close near degeneracies. Finally, two of the three visible contributions are induced by near degeneracies between electron and hole like bands. The splitting between them is induced by the superconducting ordering and is extremely small far away from the actual gap. For this reason the bands are almost touching and the Berry curvature is drastically enhanced.   
  
 \begin{figure}[h]
\includegraphics[width=\LL]{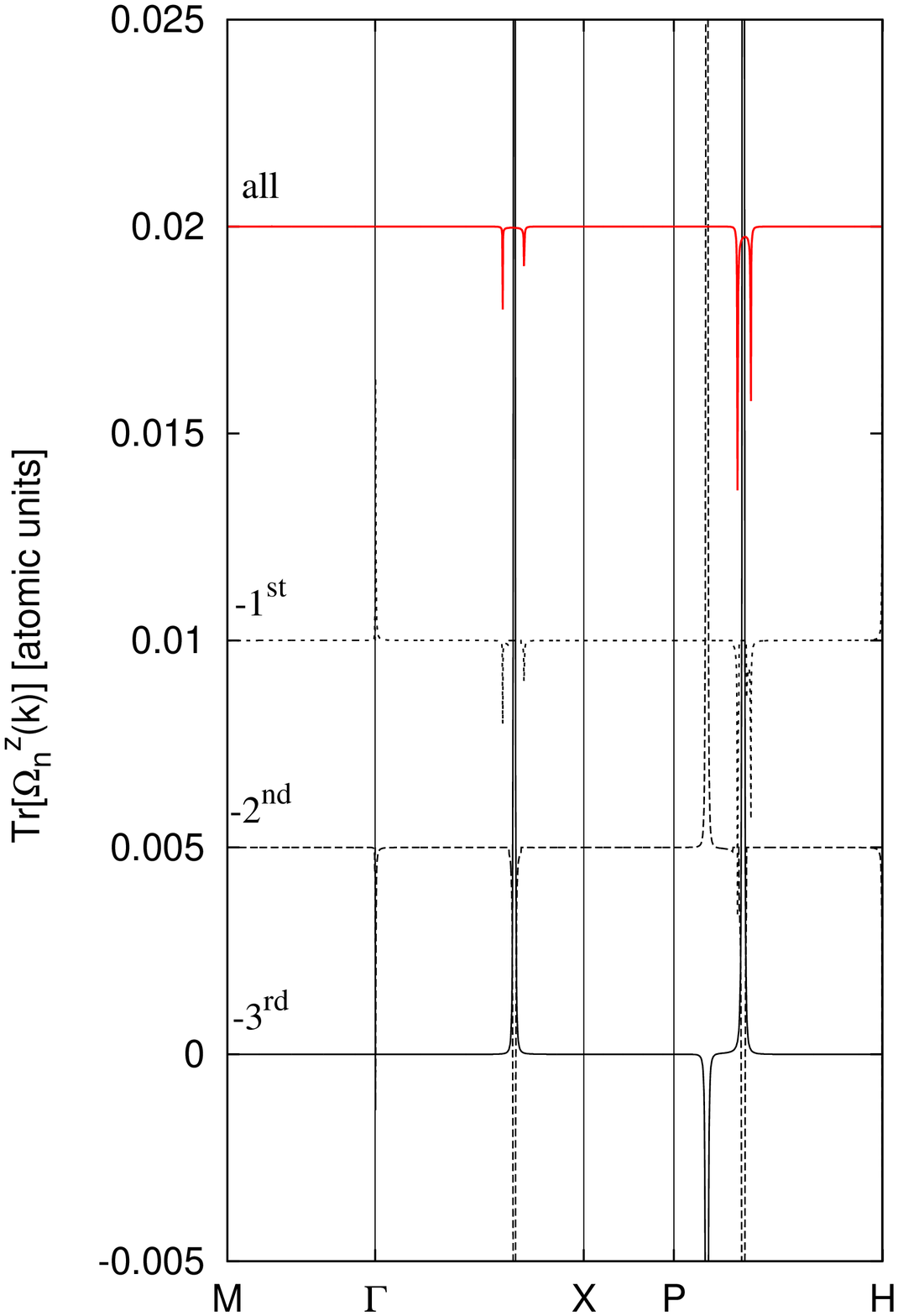}
\caption{The trace of the 2-dim. non-Abelian Berry curvature for each degenerate band in Sr$_2$RuO$_4$ along high symmetry lines without spin-orbit coupling but reducing the scale of Fig.~\ref{fig:Berry_noSOC_full} to make details visible. For visibility the curves are shifted by a constant offset and the upper curve is the sum over all bands. \label{fig:Berry_noSOC}}
\end{figure}
For Fig.~\ref{fig:Berry_noSOC} we have chosen a smaller scale to make more features visible. Evidently all contributions from the two low lying bands cancel each other in the sum and essentially only contributions from the highest band survive. Here, the curvature is induced by the near degeneracies induced by the gap between negative and positive energy states.  

As pointed out above we can connect the sum over the Berry curvature of all negative energy bands to the real part of the optical conductivity at zero frequency which was shown to be finite in the considered model. \cite{Gradhand2013} Here, we can make contact to the Berry curvature of the system and understand the effect in terms of near degeneracies induced by the time reversal symmetry breaking gap function.

With that we turn our attention to the case including spin-orbit coupling. As we discussed in section~2 a small on-site spin-orbit coupling can be included without changing any of the expressions for the Berry curvature derived above. On-site spin-orbit coupling does not induce a ${\bf k}$-dependence of the Hamiltonian, explaining why it does not contribute to the ${\bf k}$ derivative. Nevertheless, the eigenfunctions change in the presence of spin-orbit coupling, leading to new features for the Berry curvature. We considered spin-orbit coupling on the same level as in Ref.~\cite{Annett2006} and solve the BdG equation self consistently for the same set of parameters as in Ref.~\cite{Gradhand2013} including a very small spin-orbit coupling parameter of $\lambda=0.001eV$. The meaning of it is not to quantitatively describe the spin-orbit coupling within the Ru $d$-orbitals but to illustrate the effect of spin-orbit coupling in general.
 \begin{figure}[h]
\includegraphics[width=\LL]{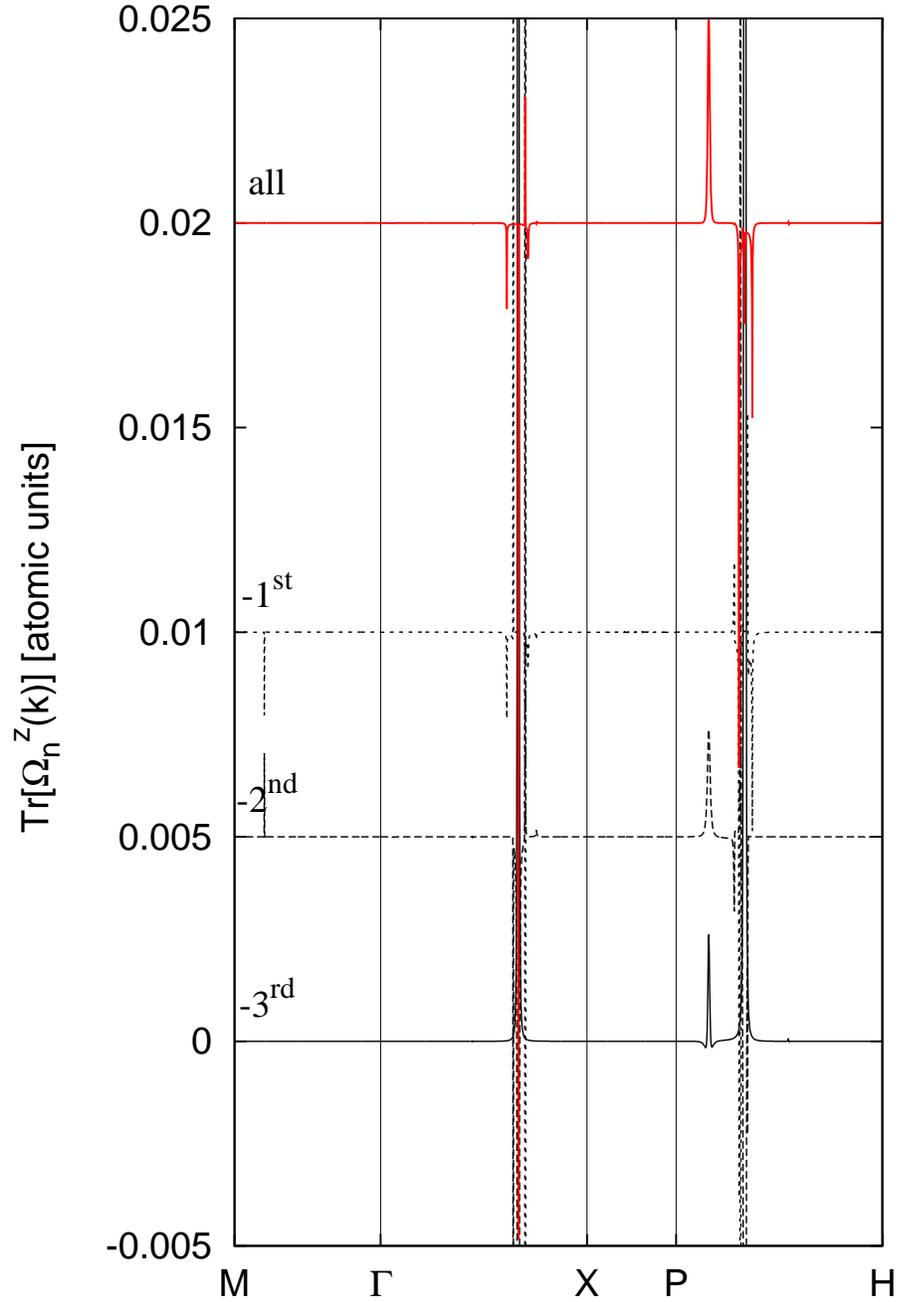}
\caption{The trace of the 2-dim. non-Abelian Berry curvature for each degenerate band in Sr$_2$RuO$_4$ along high symmetry lines including spin-orbit coupling. \label{fig:Berry_SOC}}
\end{figure}
In Fig.~\ref{fig:Berry_SOC} we show the same trace over the Berry curvature for the negative energy states as before but including a small spin-orbit coupling. Evidently new features arise which are due to the fact that by considering spin-orbit coupling we introduced a few further couplings between the orbitals along high symmetry lines. Despite this the size of the Berry curvature remains of similar order as for the case without spin-orbit coupling.
 \begin{figure}[h]
\includegraphics[width=\LL]{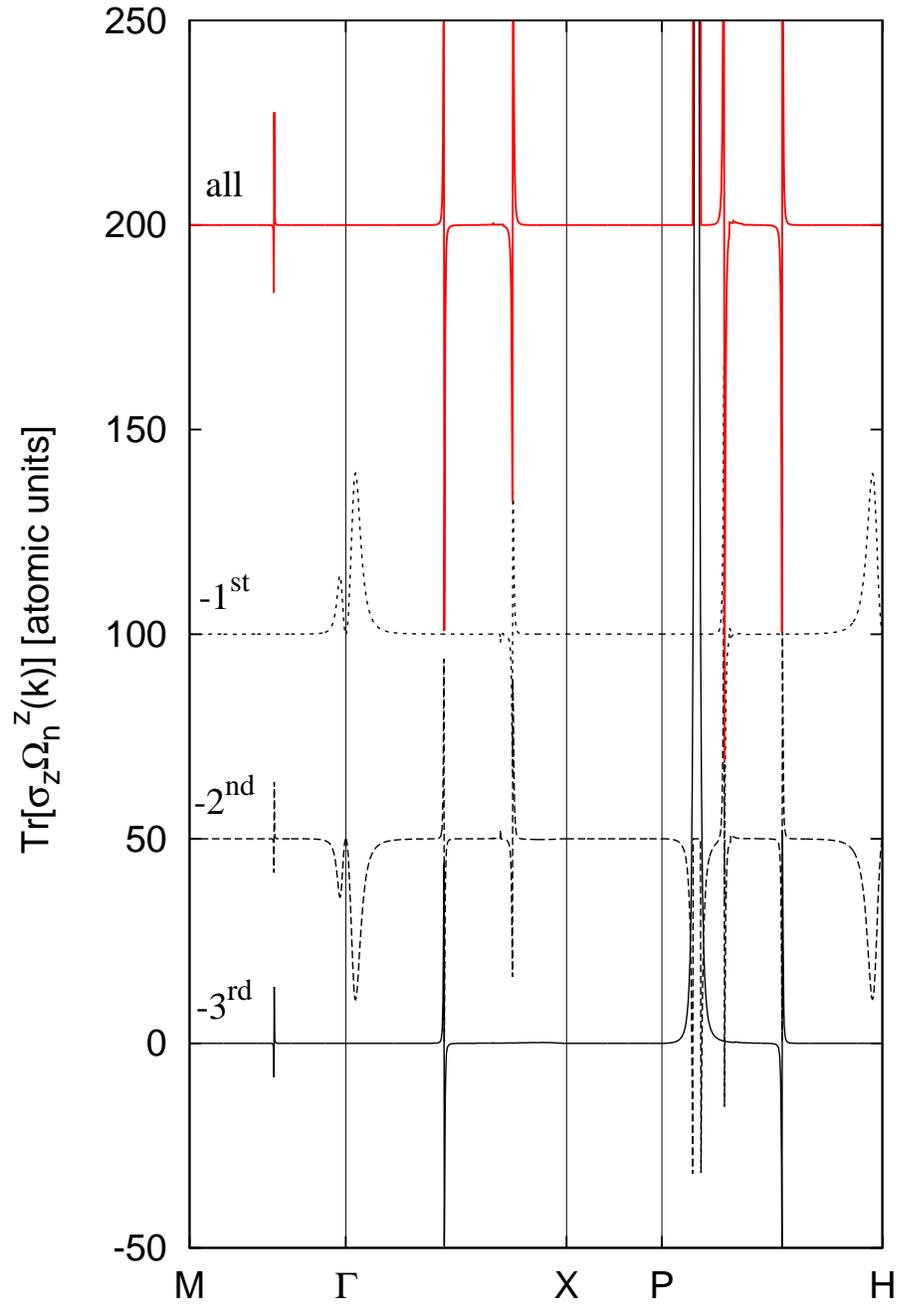}
\caption{The trace of the product of the Pauli matrix $\sigma_z$ and the 2-dim. non-Abelian Berry curvature for each degenerate band in Sr$_2$RuO$_4$ along high symmetry lines including spin-orbit coupling. \label{fig:Berry_spinTRSOC}}
\end{figure}
The picture becomes more interesting when we consider the trace over the product of the non-Abelian Berry curvature with the Pauli matrix $\sigma_z$. In the normal state it is a measure for the induced spin current or more precisely the spin Hall conductivity \cite{Gradhand2011}. This is shown in Fig~\ref{fig:Berry_spinTRSOC} and crucially this quantity is order of magnitudes larger than the ordinary trace over the Berry curvature. This quantity would be identically zero neglecting spin-orbit coupling. There is evidently a reason for that huge difference between the two traces namely the energy scale of the mechanisms inducing Berry curvatures. The gap function is at least an order of magnitude smaller than the considered spin-orbit coupling and the curvature is entirely induced by the superconducting state. In contrast, the the features of Fig.~\ref{fig:Berry_spinTRSOC} are caused by the spin-orbit coupling of the normal state Hamiltonian. While the curvature induced by the gap function would clearly vanish at $T_c$, the superconductor transition temperature, the Berry curvature induced by spin-orbit coupling would remain finite.

With that we turn our attention to the Hall sum rule as presented in the last section. To demonstrate it numerically we perform on one hand calculations for the frequency dependent imaginary part of the optical conductivity along the lines of Ref.~\cite{Gradhand2013} and perform the frequency integral from Eq.~(\ref{Eq.:sum_rule_super}). On the other hand we calculate the Berry curvature summed over the occupied bands. If we perform the ${\bf k}$ integral in addition we get only one number the real part of the Hall conductivity at zero frequency, Eq.~(\ref{Eq.:KKT}). However, we showed already in Ref.~\cite{Gradhand2013} that the Kramers-Kronig transformation holds and we hereby confirm that the Berry curvature integral gives indeed the same result in a numerically much more efficient way, as discussed for the normal state already. \cite{Wang2006} 

To give a more visual impression of the validity of the Hall sum rule we perform artificial calculations where we omit the ${\bf k}$ space integrals and present planes with constant $k_z$ with the Berry curvature summed over all occupied bands on one hand and the k-resolved imaginary part of the Hall conductivity integrated over all frequencies on the other hand. 
The result, neglecting spin-orbit coupling is shown in Fig.~(\ref{fig:Berry_sumrule}). One can see that the Hall sum rule holds nicely. In addition we show the original normal state Fermi surface lines and obviously the largest contributions come from regions where the different bands are getting close to each other. This highlights once more the importance of near degeneracies as sources 
of Berry curvatures and furthermore the driving mechanism for the optical Kerr effect in $Sr_2RuO_4$.
\newlength{\LLL} \LLL 0.5\linewidth
 \begin{figure}[h]
\includegraphics[width=\LLL]{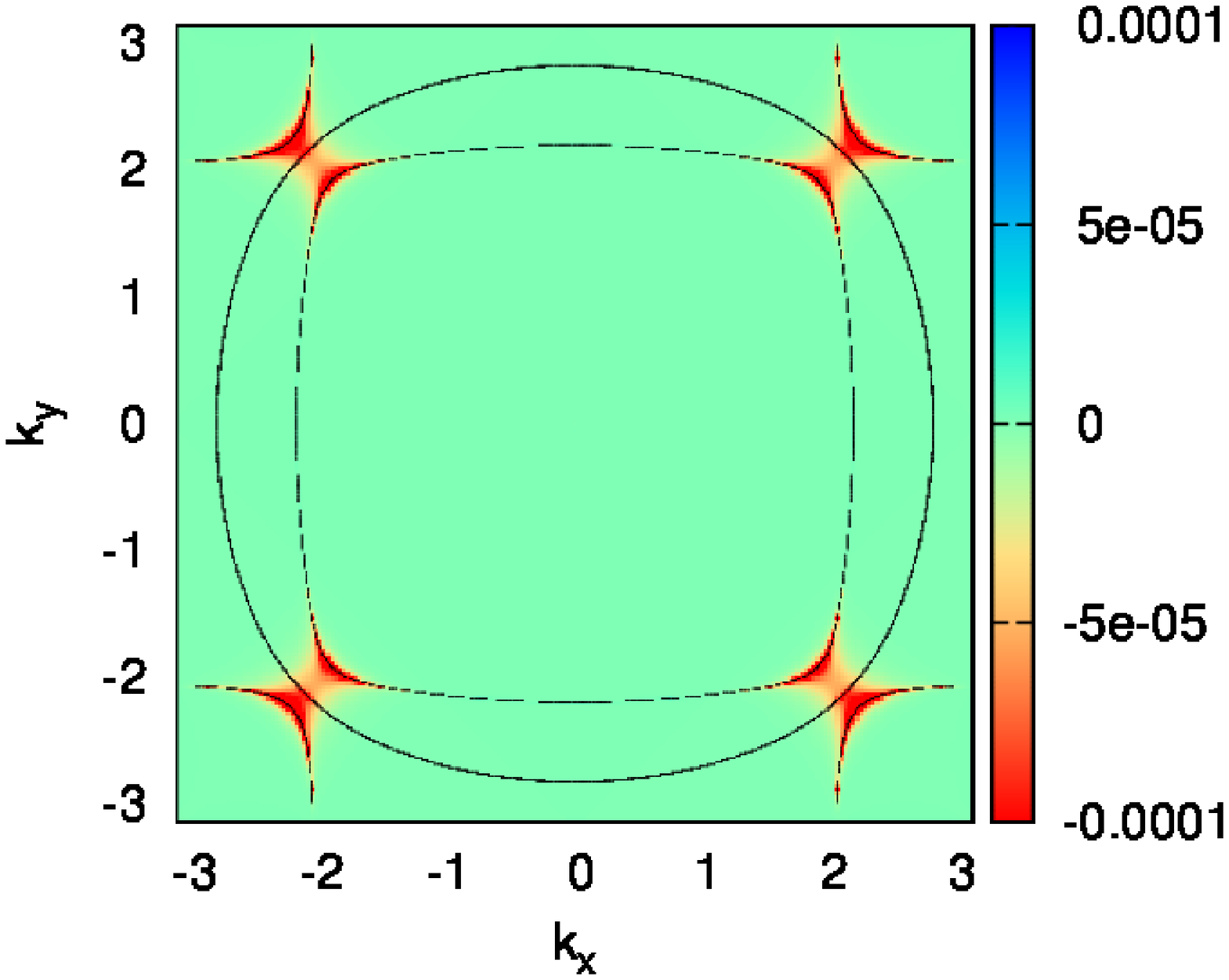}
\includegraphics[width=\LLL]{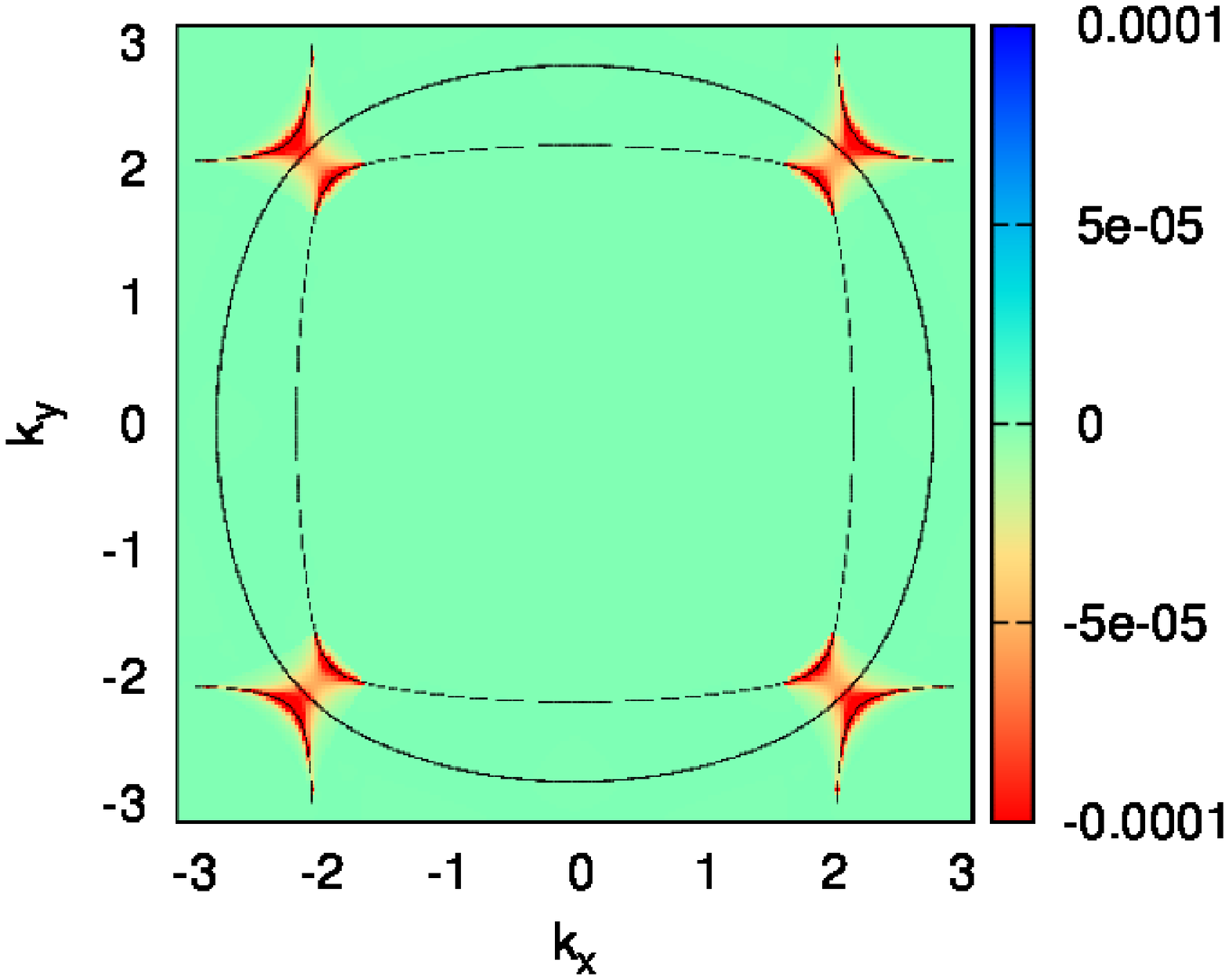}
\caption{The numerical test of the Hall sum rule established in Eq.~(\ref{Eq.:sum_rule_super}). On the left hand side the sum over the Berry curvatures for all negative eigenvalues is shown in comparison to the direct evaluation of the expression for the imaginary part of the optical conductivity according to Eq.(\ref{Eq.:Imsxy}) on the right hand site. For both figures we fixed $k_z=0.21/a$ where a is the lattice constant and the quantities in atomic units are shown as color code. The black lines indicate the normal state Fermi surface lines.  \label{fig:Berry_sumrule}}
\end{figure}
\section{Summary}
In summary we have introduced the concept of Berry curvatures to the superconducting state as described by the Bogoliubov-de Gennes equation for periodic crystals. We have shown analytically as well as numerically that the same Hall sum rule as for the normal state holds for the superconducting state connecting the Berry curvature to the imaginary part of the frequency dependent optical conductivity. We applied the methodology to the unconventional superconductor $Sr_2RuO_4$ highlighting the different features induced by the time reversal symmetry breaking of the gap as well as the spin-orbit coupling. This allows to get a deeper insight into the mechanism leading to the intrinsic optical Kerr effect in the superconducting state of $Sr_2RuO_4$ which is entirely a multiband effect vanishing for single band models.

\end{document}